# 3D Printed Multilayer Structures for High Numerical Aperture Achromatic Metalenses

Cheng-Feng Pan[1,2], Hao Wang[1,3,4,*], Hongtao Wang[1,2], Parvathi Nair S[1,5], Qifeng Ruan[6], Simon Wredh[1], Yujie Ke[5], John You En Chan[1], Wang Zhang[1], Cheng-Wei Qiu[2], Joel K. W. Yang[1,*]

Affiliations

[1] Engineering Product Development, Singapore University of Technology and Design, Singapore 487372, Singapore

[2] Department of Electrical and Computer Engineering, National University of Singapore, 4 Engineering Drive 3, Singapore, 117576, Singapore

[3] College of Mechanical and Vehicle Engineering, Hunan University, Changsha 410082, China

[4] Greater Bay Area Institute for Innovation, Hunan University, Guangzhou 511300, China

[5] Institute of Materials Research and Engineering, A*STAR (Agency for Science Technology and Research), Singapore 138634, Singapore

[6] Ministry of Industry and Information Technology Key Lab of Micro-Nano Optoelectronic Information System, Harbin Institute of Technology (Shenzhen), Shenzhen 518055, China

Corresponding author

✶E-mail: joel_yang@sutd.edu.sg

✶E-mail: hao_wang@sutd.edu.sg

## Abstract

Flat optics consisting of nanostructures of high-refractive-index materials produce lenses with thin form factors that tend to operate only at specific wavelengths. Recent attempts to achieve achromatic lenses uncover a trade-off between the numerical aperture (NA) and bandwidth, which limits performance. Here we propose a new approach to design high NA, broadband and polarization-insensitive multilayer achromatic metalenses (MAM). We combine topology optimization and full wave simulations to inversely design MAMs and fabricate the structures in low-refractive-index materials by two-photon polymerization lithography. MAMs measuring 20 μm in diameter operating in the visible range of 400–800 nm with 0.5 NA and 0.7 NA were achieved with efficiencies of up to 42%. We demonstrate broadband imaging performance of the fabricated MAM under white light, and RGB narrowband illuminations. These results highlight the potential of the 3D printed multilayer structures for realizing broadband and multi-functional meta-devices with inverse design.

## Introduction

Recent progress in metalenses at micro and macro length scales have shown unprecedented potential to achieve remarkable imaging performance, finding applications in areas such as light-field imaging, bioanalysis, medicine, semiconductor, and quantum technologies. In particular, achromatic lenses exhibit broadband responses that enable capture of colour information, which expands design possibilities and expands the application scenarios for photonic devices[1, 2, 3, 4, 5, 6, 7, 8, 9]. The desirable characteristics, e.g. ultra-compact, ultra-thin and light-weight characteristics, make metalenses compelling in imaging systems[10, 11, 12, 13, 14, 15, 16]. However, most metalenses are patterned in high-refractive-index (HRI) materials that provide good optical control due to strong light confinement but are highly dispersive, making broadband implementations challenging.

The Abbe number is a figure of merit in lens design that represents how dispersion free a transparent material is, i.e., higher values indicate lower dispersion. Commonly used HRI materials in metalenses such as Si, $TiO_2$ and GaN have Abbe numbers of 5.5, 9.86 and 18.7, respectively. In contrast, the Abbe number of a typical resin IP-Dip with refractive index ~1.56 is ~35.2 (details in Supplementary Note 1). As HRI materials are highly dispersive and the sub-wavelength nanoantennae also induce strong scattering in such meta-devices, the concomitant realization of metalenses with (1) high numerical aperture (NA), (2) broadband (achromatic), and (3) high efficiency in the visible range is hindered. Some of the recent works on achromatic metalenses tackle these three points in the near infrared or mid infrared range, where the material dispersion is significantly lower and material absorption is negligible[17, 18, 19].

Single lenses based on traditional refractive optics and diffractive optics exhibit chromatic aberrations, as shown in Fig. 1b. Single layer multi-level diffractive lenses (MDL)[20] and metalenses minimize chromatic aberration at the expense of low NA, due to the intrinsic trade-off between bandwidth and NA for a given lens radius[11],

$$R_{max} \leq \frac{C}{\Delta\omega\left(\frac{1}{NA} - \sqrt{\frac{1}{NA^2} - 1}\right)} \qquad (1)$$

where $R_{max}$ is the maximum achievable radius of the lens, C is a constant related to the material, central frequency, and structure, $\Delta\omega$ is the bandwidth. The premise of Eq. (1) is that the phase distribution of the generated wavefront meets the requirement of an aberration free metalens[21]. Therefore, this formula is the relationship that needs to be satisfied to realize a high efficiency focusing lens. It can be further derived that for a given lens radius, the maximum bandwidth is inversely proportional to NA, i.e., $\Delta\omega \lesssim \frac{C_1}{NA}$. $C_1$ is also a constant related to the material, central frequency, and structure.

Multilayer metalenses are not constrained by Eq. (1), which holds only for the case of single layer metalenses. In conventional refractive lenses, the lens doublet, is often used to reduce chromatic aberrations, which uses two different materials and geometric designs to compensate the dispersive response. Similarly, multilayer metalenses have been explored to introduce additional degrees of freedom (DOF)[22, 23, 24, 25], because of the constraints in simultaneously improving NA and bandwidth in single layer metalenses. With the additional DOF available for the design of the multilayer structures, enhanced abilities of light manipulation by wavefront control are made possible. Indeed, by using synergistic effects between the individual layers, the multilayer structure can be designed to break the design trade-off between high NA and bandwidth. (Fig. 1c).

The fabrication challenges associated with producing MAMs is lifted using nanoscale 3D printing. Due to the high-resolution patterning requirements, electron beam lithography (EBL) has been widely used to fabricate nano-scale meta-devices including metalenses. However, patterning multilayer and variable height structures using EBL is inconceivable, requiring alignment and multi-step fabrication processes. Conversely, nanoscale three-dimensional (3D) printing enables the patterning of a multilayer metalens in one lithographic step, which facilitates rapid prototyping of complicated structures. With two-photon polymerization lithography (TPL), various 3D designs have been realized, e.g., complex micro lenses[8, 26, 27, 28], gradient index lenses[29], diffractive lenses[30], and other wavefront modulation elements working in the visible band [31, 32]. The crosslinked resin in TPL are polymeric acrylates with low refractive index (LRI) ~1.5, negligible absorption, with an Abbe number of ~35.2 (for the IP-Dip resin used here)[33]. Hence, the sensitivity to changes in wavelength is smaller compared to the more commonly used HRI materials in metalens design. Thus, TPL patterning with LRI polymers mitigates the material dispersion problem and enables high-resolution fabrication of complex 3D structures required in MAMs. Its ability to pattern sub-wavelength features to modulate the wavefront have been demonstrated with applications in the visible and IR range[34, 35, 36, 37].

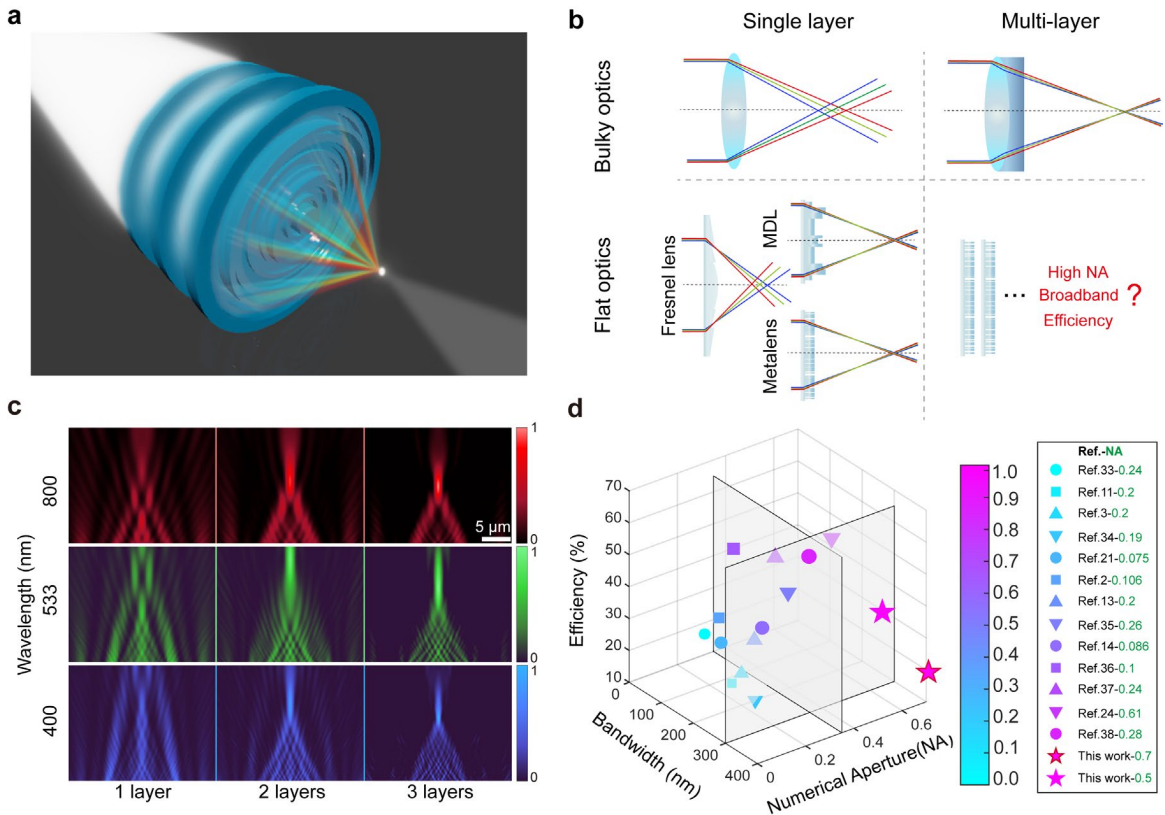

**Fig 1: Achromatic performance of different single lenses and multilayer lenses.**

**a** Schematic of the designer 3D printed multilayer achromatic metalens (MAMs). **b** Schematic of traditional and flat optics lenses (including Fresnel lens, multilevel diffractive lens (MDL) and metalens) with single and multilayers. **c** Evolution of focal spots as additional layers are added at wavelengths of 400, 533, and 800 nm (results are from an optimized 3-layer design of 0.5 NA MAM). **d** The efficiencies, numerical apertures, and bandwidths (works in the visible band) of various achromatic metalenses[2, 3, 10, 12, 13, 22, 25, 38, 39, 40, 41, 42, 43] (Details in Supplementary Note 2). The colour bar and marker size represent the figure or merit defined as the square root of the sum of squares of efficiency, NA, and bandwidth. The grey planes indicate previous limits at bandwidth = 300 nm and NA = 0.35. The NA values of each metalens are shown in the legend.

In this work, we utilized topology optimization (TO) with the adjoint method to obtain suitable design parameters to achieve achromatic lensing behaviour. This approach enables inverse design of nanostructure heights within each layer at varying distances between layers and number of layers to achieve high efficiency metalens. By using the optimized geometric parameters and the inverse designed structure, we tailored our fabrication process to achieve a stable, multi-layer and high-resolution structures in a short fabrication time of ~1 min for each MAM. Employing shrink-compensation and proximity-effect reduction techniques, we successfully 3D printed the multilayer achromatic metalens (MAM, schematic in Fig. 1a) with 400 nm bandwidth ranging from 400 nm to 800 nm, NA of 0.5 and 0.7. We discussed the effect of edge rounding and smoothing due to limited resolution during 3D printing and their influence on the MAM performance. Simulation results agree with the measured achromatic performance. Lastly, we demonstrated the performance of the MAM by imaging a resolution target with white light illumination. The 0.5 and 0.7 NA MAMs achieve an average focusing efficiency of 42% and 20%, respectively. Here, the focusing efficiency is defined as the ratio of the power at the focal spot within the circle with a diameter of 3×FWHM (full width at half maximum) to the total incident power on the MAM[44].

The MAMs demonstrate previously unachievable performances in efficiency, bandwidth and NA as shown in Fig. 1d, in which the figure of merit is defined as the square root of the sum of the squares of three parameters. Compared with reported achromatic metalenses, the MAMs overcame limits in bandwidth, and NA, as shown in Fig. 1d. Our work integrates the advantages of LRI materials, nanoscale high-resolution 3D printing, and topology optimization to achieve MAMs with exceptional performance, which may inspire a new paradigm for the design and fabrication of multi-functional broadband optical elements and devices.

## Results

**MAM Design.** The principal difference between a multi-level metalens and a multi-level diffractive lens[45] is the size of the smallest feature. In this work, the minimum feature size is designed to be 0.2 μm, the 0.2 μm width is set by the resolution limit of TPL achievable in our lab with our equipment. Thus, it requires full wave simulations to account for interlayer interaction and scattering. The proposed model is shown in Fig. 2a for the diametric cross-section of the circular MAM. The darker regions in Fig. 2a are the solid parts of the MAM consisting of crosslinked IP-Dip resin. In the model, $R = 10$ μm is the lens radius with a 2 μm thick ring used as structural support. The designed focal length $f = \frac{R}{\tan(\sin^{-1} NA)}$. The thickness of the supporting layer $h_s = 0.5$ μm, and $h_i$ is the distance from the bottom of the $n^{th}$ supporting layer to the top of $(n-1)^{th}$ layer beneath it. To capture most of the light from one layer to the next, i.e., minimizing losses caused by radial diffraction, $h_i$ is set to be less than 2 μm (Supplementary Note 3). $\Omega_n$ is the design region consisting of n rectangular regions measuring 20 μm wide and 1.2 μm tall, one for each corresponding layer. In $\Omega_n$, we defined 100 rectangles with 0.2 μm width, each corresponding to the cross section of a ring around the lens. Due to axial symmetry, we only need to optimize for the heights of 50 rectangles. The maximum height of each ring is chosen to be 1.2 μm according to simulations showing that rings of heights 0–1.2 μm high will achieve the requisite geometric phase of 0–2π. The height of the rectangle was discretized to 31 levels (30 blocks along the z axis). In design region $\Omega$, a tensor $\xi$ (n, 50, 31-1) can be used to represent all the variables. Their values range from 0 to 1, where '0' corresponding to air and '1' corresponding to resin. Other values will use the effective refractive index to represent. In addition, we take filtering and binarization steps before we turn the designed structure into a real structure. We varied $h_i$ from 0 to 2 μm and number of layers from 2 to $n$, respectively in each optimization round. In the $p^{th}$ round, we optimized and obtained the variable tensor $\xi_p$ according to $h_{ip}$ and $n_p$.

For a single wavelength, the optimization goal in designing a focusing metalens is to maximize the intensity $\Phi = |E(\xi)|^2$ at the focal spot of the lens at the specific wavelength. While for a broadband case, where the goal is to increase the overall focal intensity, the strategy is to maximize the minimum focal intensity across all wavelengths. Thus we have the following function[41]

$$\max\left(\min(\Phi(\xi, \omega))\right) = \max\left(\min(|E(\xi, \omega)|^2)\right) \qquad (2)$$

**Fig 2: Topology optimization of the MAM with different layer number and spacing distance.**

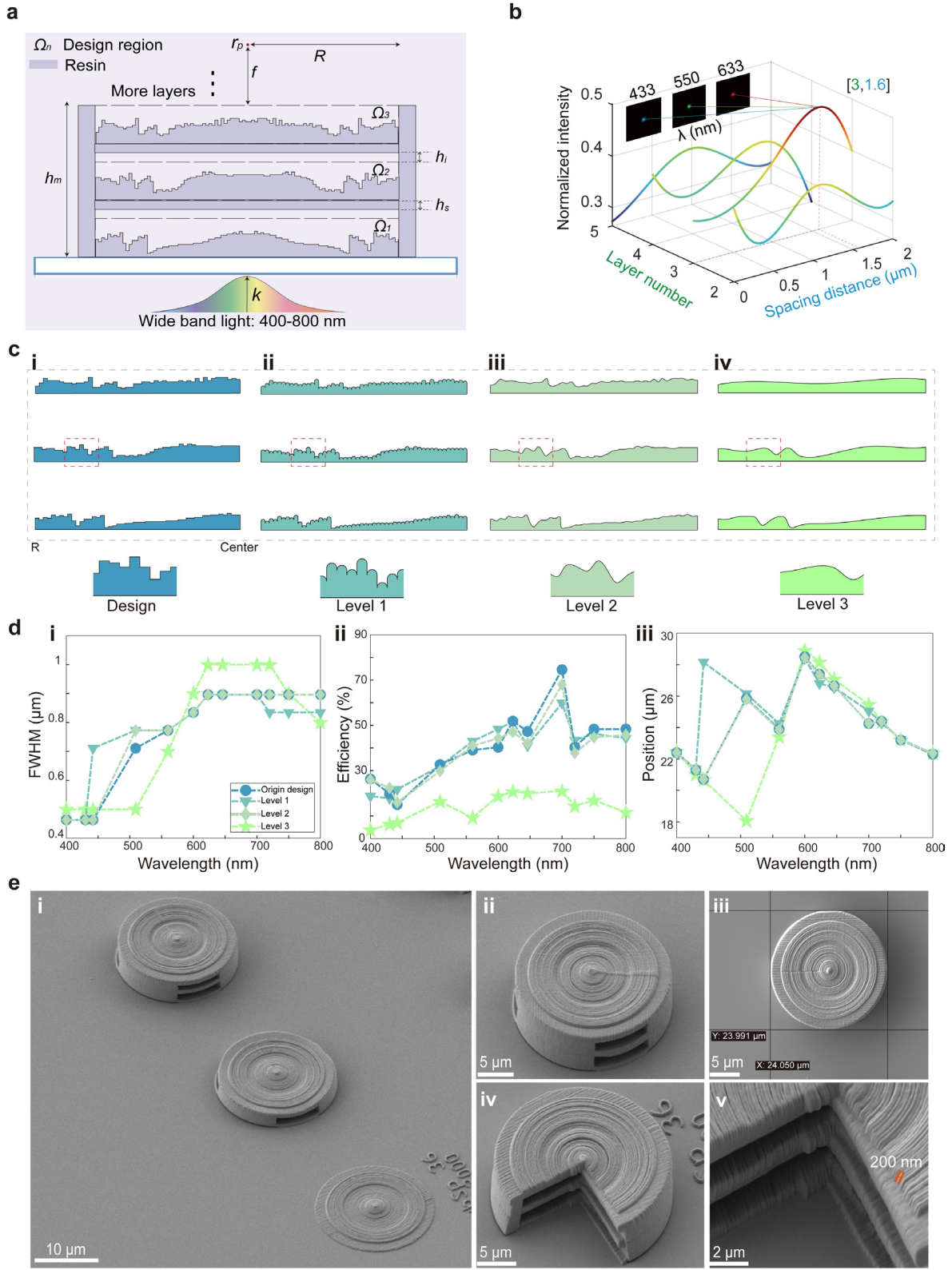

**a** Design model and schematic of the optimization region with indicated parameters described in the text. **b** Relations of the normalized intensity with the layer number and spacing distance. With the inverse design, the best case is located at [$l$, $sp$] = [3, 1.6 μm]. **c** Schematic of the edge rounding and surface smoothness approximations at different levels, initial design (i), level 1 by rounding the top (ii). Level 2 is generated by applying 10 nm relative tolerance interpolation to the origin height vector (iii), and level 3 is generated by applying 20 nm relative interpolation. (iv). **d** Calculated FWHM (i), efficiency (ii) and position of maximum focal intensity along the propagation axis (iii) for different levels. The efficiency (ii) is calculated at the focal plane corresponding to the maximum focal intensity. **e** Tilted view SEM images of the fabricated MAM with 0.5 NA: (i) Deconstructed MAM showing single, double, and triple (full) layers; (ii) Enlarged view of full MAM, (iii) top view and size of the MAM, (iv-v) Sectioned MAM revealing internal structure and details of the 200-nm wide ring structures.

Here, the topology optimization with the assist of adjoint method is applied on the material distribution and geometric configuration to achieve the convergence of the functions above. TO algorithm optimizes the material layout within the user-defined space for a given set of conditions and constraints. To solve the multilayer topology optimization problem, the parameters for all layers should be included in one integrated tensor $\xi$ to be optimised simultaneously[46, 47, 48]. Thus, the broadband metalens optimization goal can be written as

$$\max_{\xi(r)}(\min(\Phi(\boldsymbol{\xi}, \boldsymbol{\omega}))) \qquad (3)$$

$$s.t. \quad \varepsilon_r(\boldsymbol{\xi}) = \varepsilon_{r,IP-Dip} + \boldsymbol{\xi} \cdot (\varepsilon_{r,IP-Dip} - \varepsilon_{r,Air}) \qquad (4)$$

$$0 \leq \boldsymbol{\xi} \leq 1 \qquad (5)$$

Here, $\varepsilon_r(\xi)$ is the position-dependent effective permittivity profile used in the inverse design process. The full inverse design process is illustrated in Supplementary Note 4. As mentioned before, in each iteration we optimized and obtain the variable tensor $\xi$ for different $h_i$ and $n$. With additional layers beyond a certain point, we expect the overall efficiency to decrease, due to multiple interfacial reflections and scattering. In addition, with smaller spacing distance, the stronger interaction between layers tends to reduce efficiency, while larger distances result in high-order diffraction and more energy losses. Note that the optimized performance for the single layer metalens was poor, as expected from the single layer limitation between NA and bandwidth (results in Supplementary Note 5). Thus, as shown in Fig.2b, we searched through the parametric space with layer number $n$ from 2 to 6 and spacing distance $h_i$ from 0 to 2 μm (0, 0.25, 0.5, 1.0, 1.5, 2.0, 2.5) initially for MAM with 0.5 NA. The efficiency starts to decrease after 3 layers and drops dramatically when the number of layers reaches 6, due to the diffraction away from the optical axis between layers.

We found an initial optimal performance at 3 layers and a spacing distance in the range from 1 to 2 μm. Setting $n = 3$, we ran another round of optimization for spacing distance $h_i$ in the range of 1 to 2 μm with a step of 0.1 μm. The final design has 3 layers and a spacing distance of 1.6 μm, which shows a peak in normalized intensity from the approximate fitted parametric exploration curves in Fig. 2b. Though the above process was optimized for MAM with 0.5 NA, it is also applicable for the 0.7 NA case to achieve an optimal achromatic design (Details in Supplementary Note 6).

**Influence of print resolution.** The samples were fabricated using the Nanoscribe GmbH Photonic Professional GT2, a TPL based 3D printing system utilizing a galvo-scanned focused beam from a 780 nm femtosecond laser to induce crosslinking of a liquid resin into a nanoscale solid voxel at the focal spot (details in Methods). One voxel is an ellipsoid that takes the shape of squared intensity profile of the focused Gaussian beam of the laser[49]. As we printed at the resolution limit of the TPL, the fabricated structures will inevitably exhibit deviations from the nominal design of Fig. 2c (i). Studying these deviations, such as corner rounding and loss of details, can inform us about the fabrication process and improve our design approach. In Fig. 2c, we introduced corner rounding by introducing a fillet to the top edges while preserving the structure heights, resulting in Fig. 2c (ii). To introduce a more severe effect of variation on the surface profile, we adopted the linear interpolation to (iii) and (iv) with different relative tolerance, and the relative tolerance means the maximum allowed distance between the points in the generated height curve and the corresponding points in the sequence of heights we input. We interpolated the original design (height curves) with a relative tolerance of 10 nm, generating the structure depicted in Fig. 2c (iii). The topography is smoother while retaining most details. Similarly, using interpolation with a tolerance of 20 nm, we obtained Fig. 2c (iv). Within Fig. 2c, we observed the disappearance of sharp edges from (i) to (ii). From (iii) to (iv), only the overall surface morphology is preserved, with significant loss of details. We conducted simulations of the corresponding MAMs for the four different structures and compared their performance. As shown in Fig. 2d (i)-(iii), the efficiency and FWHM are almost unchanged for levels 1 and 2 compared to the original design. But for level 3, where most of the height information is lost, we observed a drastic drop in the efficiency. We extracted the focal point position variation on the propagation axis as shown in Fig. 2d (iii). Except for 2 data points at 442 nm of level 1 and at 509 nm of level 3, the 4 different shapes have good consistency in their ability to achieve broadband focusing of light. The reason for the 2 outliers is the presence of two focal points on the axis of light propagation, and the selection of the focal position based on higher intensity (In Supplementary Note 7). In Fig. 2d (iii), longer wavelength shows good consistency due to the less sensitivity to small features. If we consider the balance of manufacturing feasibility and performance (we also compared the point spread function and modulation transfer function (MTF) of the original structure and level 1 structure which is included in Supplementary Note 8. Original structure and level 1 structure show good consistency on the MTF performances), original design and level 1 are the better choices. In the next section, based on this result, we optimized the fabrication method to achieve a prototype close to the nominal design.

**Experimental results.** With the optimized parameters, we experimentally demonstrated MAMs with NA at 0.5 and 0.7, possessing high focusing efficiency with broadband achromatic performance. By using TPL, we precisely realized the designed MAM with nanoscale feature sizes. As the width of the ring measured only 200 nm, overcoming proximity effects during printing of the dense rings becomes a challenge, especially in in the central region, which would be overexposed and end up being taller than designed. To circumvent this effect, we introduced a 500 ms interval between the exposure of adjacent rings and exposed the rings in sequence starting from the centre and moving towards edge. A detailed study of this procedure is beyond the scope of this work. In addition, we optimized the interval of printing in the $z$ direction for different height structure, further improving the precision. For each layer, the outer circular support was first printed with rings, then the support layer was printed with straight bridging lines to increase mechanical stability. Several windows were also added on the outer support to allow drainage of unexposed resin during development. Because of the limited degree of crosslinking of resin during exposure, 3D printed structures usually suffer from shrinkage. Thus, in addition to using post development ultraviolet exposure to increase the degree of curing, we also compensated for shrinkage by increasing the nominal size each layer to achieve the intended size of the design. We first printed the structures without compensation to determine the volumetric shrinkage from scanning electron microscopy (SEM) inspection. Afterwards, successful MAMs were obtained in the final fabrication (Fig. 2e (i)-(v)) with compensation added by scaling the 3 layers by a factor of 1.01,1.02 and 1.03 from bottom to the top (details in Supplementary Note 9). We also printed a separate structure with a section removed (Fig. 2d (iv-v)) to verify the dimensional agreement with the design and observe the good separation between individual rings and layers within the MAM.

**Fig. 3: Characterization of achromatic performance of the fabricated MAMs.**

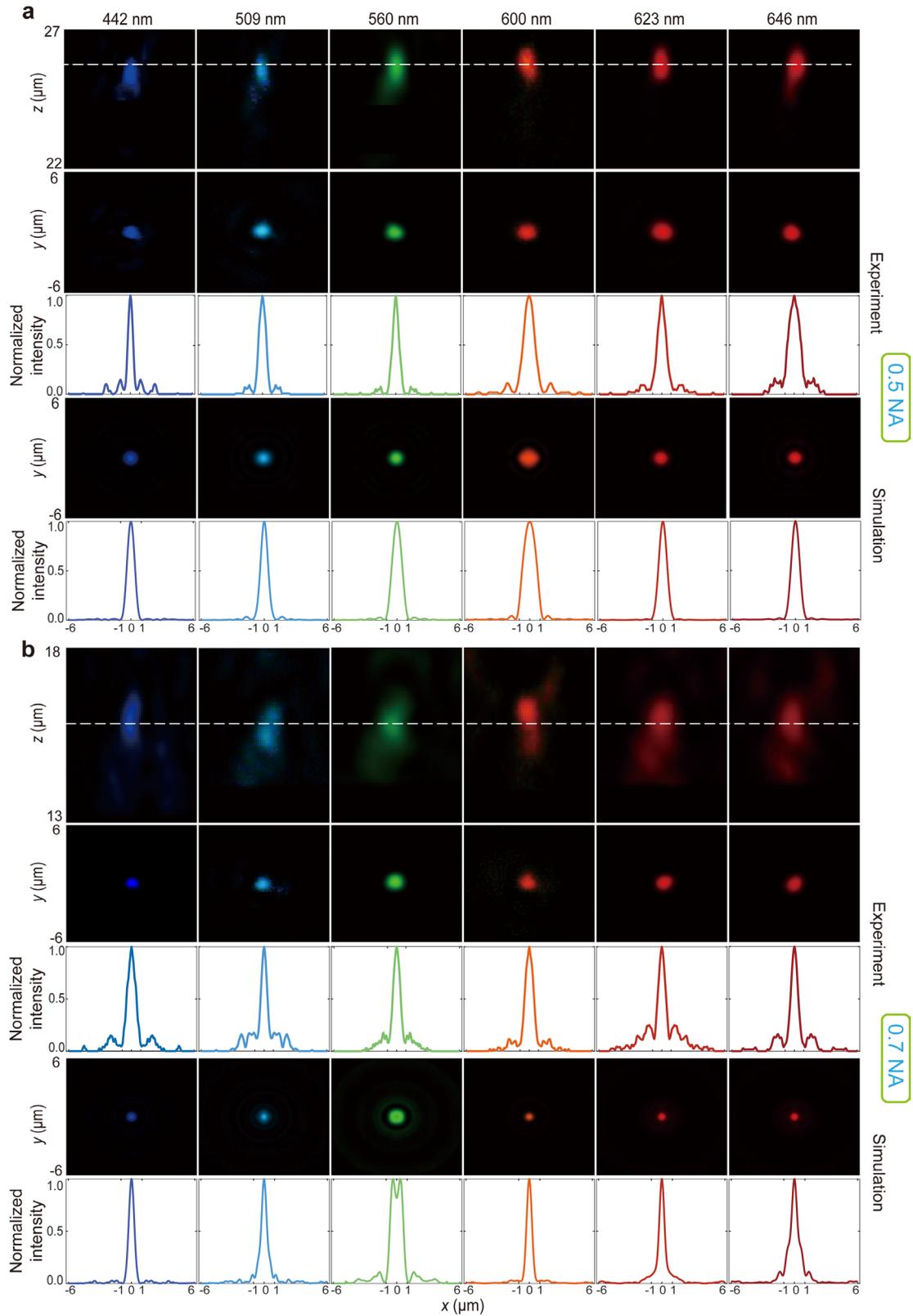

Intensity distributions on axial and focal planes for MAMs with **a** 0.5 NA and **b** 0.7 NA. In both panels, the first row is the experimentally measured normalized focal length shifts in the *xz* plane, *z* axis distance measured from the top of the sample, the second row is the measured intensity distributions in the *xy* plane, and the third row is the normalized intensity along the white dashed lines at the focal plane. The fourth and fifth rows are the corresponding simulation results (details in Supplementary Note 6).

We characterized the achromatic focusing performance of the MAMs by measuring the distribution of light intensity in the focal plane (*xy*) and along the propagation plane (*xz*) with a supercontinuum laser source. As seen in Fig. 3a-b, the lengths of the focal spots at different wavelengths are less than 2 µm, and the focal positions shift by less than 3 µm across the visible spectrum. We observed good agreement between simulated and measured results for the intensity distributions at the focal and axial planes for both 0.5 NA and 0.7 NA MAMs at six discrete wavelengths of 442, 509, 560, 600, 623, 646 nm, as shown in Fig. 3. Measurements beyond the range of 442–646 nm were not conducted due to limitations of our CMOS sensor. The experimental focusing efficiency across the wavelengths ranges in 22.5–42.7% for 0.5 NA, with corresponding simulated efficiencies ranging from 24.7% to 74.2%. For the 0.7 NA, the experimental focusing efficiency across the wavelengths is in the range of 13.3–22.1%, appearing to closely match the corresponding simulated efficiencies of 14.6–26.1% (Fig. 4a and b). The inconsistency between measurements and simulation is likely due to imperfections in the fabricated lenses as explained previously, a non-ideal collimated light source, and stray light. The wavelength averaged simulated focusing efficiency is 44.2% and 20.0%, for 0.5 and 0.7 NA, respectively. The simulated FWHM is ~0.88–1.4 µm for 0.5 NA MAM and 0.64–1.8 µm for 0.7 NA MAM, respectively.

We assessed the imaging quality of the MAM with NA of 0.5 by placing it onto a USAF 1951 resolution target with ~100 µm spacing between the top of the MAM and the top of the resolution target. A minimized image of group 6 element 3 in the resolution target was imaged through the MAM which has a spacing distance of 3 times the focal length to the objectives[15]. The images were captured under a microscope under white light illumination and with colour filters at wavelengths of 450 nm, 532 nm and 633 nm placed in front of the white light source as shown in Fig. 4c. The results demonstrated that the fabricated MAMs performs well under incoherent white light for achromatic imaging applications (More images can be found in Supplementary Note 10). We also fabricated a 0.5 NA spherical lens and a Fresnel lens for comparison to demonstrate the unrivalled ability of the MAMs in removing chromatic aberration (see Supplementary Note 11).

**Fig. 4: Focusing efficiency and imaging performance of MAM.**

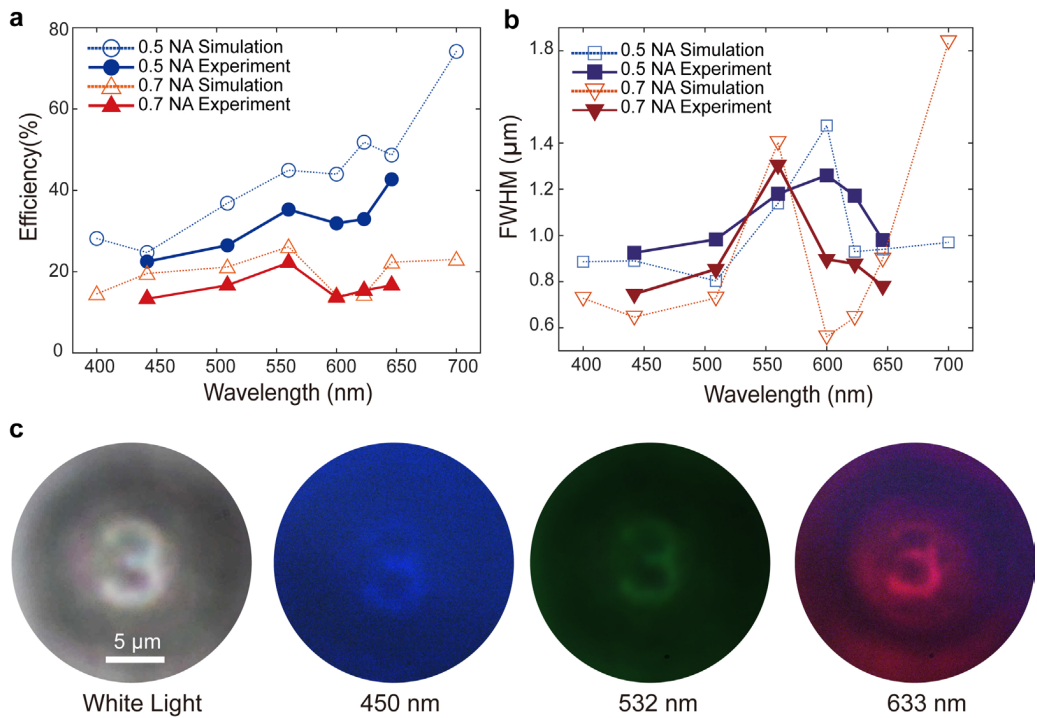

**a** Comparison of experiment and simulated broadband focusing efficiencies for MAMs with NA of 0.5 and 0.7 at the same focal plane defined by NA. **b** Comparison of experiment and simulated broadband FWHM for MAMs with NA of 0.5 and 0.7 at the same focal plane defined by NA. **c** Optical images of the number "3" in group 6 element 3 in the USAF 1951 resolution target captured through the 0.5 NA MAM under white light and applied blue (450 nm), green (532 nm), red (633 nm) filters.

## Discussion

In a multilayer metalens system, each layer can be regarded as an achromatic corrector and focusing element. The MAM overcomes constraints that trade-off between NA and bandwidth in single layer achromatic metalens. While three layers enables high efficiency, MAMs with higher NA can be achieved by adding layers to the system at the expense of efficiency. The effect of spacing distance on the interaction between layers is difficult to quantify with a physical model, and thus efficient and accurate inverse design methods, like the topology optimization presented herein, are particularly useful to design high performance structures. We found that for distances smaller than 0.5 μm, interlayer coupling greatly reduces the transmitted power. Conversely, if the spacing distance is larger than 2 μm, high-order diffraction losses also decrease the efficiency. Hence, we performed a high-resolution search within these limits to find an optimal spacing distance of 1.6 μm. Our fabricated MAMs with broadband operation in the wavelength regime of 400–800 nm and average focusing efficiencies of ~44.2% for 0.5 NA and 20% for 0.7 NA demonstrates the potential of the inverse design and TPL fabrication methodology. This work pushes the boundary of metalens design to the elusive region where high NA, broadband operation, and high achromatic efficiency can be achieved simultaneously.

Our results demonstrated that stacked metasurfaces based on LRI materials overcomes the limitations of single-layer flat optics to extend the performance of metalenses into broadband operation while preserving high NA. We harnessed the power of topology optimization and adjoint method to accelerate the iterative design of metalenses, and used it to overcome the design constraint of good achromatic performance with simultaneously high NA. The presented design framework, and experimental fabrication by TPL at a resolution of ~200 nm provides a practicable path from 2D metasurfaces to complex 3D metasystems. In the future, with higher resolution 3D printing techniques and HRI resins, we envisage an increase in multi-functional optical systems that operate with a broadband response beyond the visible range.

## Methods

## Finite-difference time domain (FDTD) simulations

The full wave simulations of the MAMs were carried out with Lumerical FDTD and the adjoint method algorithm was conducted by the Lumerical script. In the adjoint method process, a 2D cross section of the MAM in the *xz* plane was chosen for the simulation model as shown in Fig. 2b. The boundary conditions are set as PML. The light source constituted of both *x* and *y* polarized light sources with the same amplitude to represent unpolarized light. The TO algorithm was conducted with an in-house MATLAB code.

## Fabrication with two-photon polymerization lithography

The fabrication of MAM was performed using a two-photon polymerization lithography-based 3D printing technique (Nanoscribe GmbH Photonic Professional GT2). A 63× objective lens with NA = 1.4 was used in dip-in laser lithography mode. For the fabrication of the supporting layer and supporting pillars, ContinuousMode in GalvoScanMode was applied with the PiezoSettlingTime of 50 ms and GalvoSettlingTime of 4 ms. For the fabrication of the metalens structures, the PiezoSettlingTime was changed to 500 ms and GalvoSettlingTime remained as 4 ms. The printing parameters for the supporting layer and supporting walls were ScanSpeed 6,000 μm/s, LaserPower 40% of the total power (50 mW). While for the centre part of the MAM structures (from the centre to the 10$^{th}$ rings), ScanSpeed was 60,000 μm/s, and LaserPower was 30%, corresponding to 15 mW. For the left MAM structures (from the 11$^{th}$ to the 50$^{th}$ rings), ScanSpeed was 40,000 μm/s, and LaserPower was 32%, corresponding to 16 mW. The waiting time between printing two rings was set as 500 ms. After the exposure process, the sample was developed in propylene glycol monomethyl ether aceta for 6 min, isopropyl alcohol for 5 min, and methoxy nonafluorobutane for 7 min. When emerged in isopropyl alcohol, the sample was cured with UV light for 5 min with 70% maximum power (DYMAX, MX-150, 405 nm).

## Characterization

A supercontinuum light source (SuperK SELECT UV/nIR1, NKT Photonics) was used for the characterization of achromatic performance within the continuous bandwidth of 450–650 nm. The optical testing apparatus is included in Supplementary Note 12. The laser passed through a pair of periscopes and collimated along the optical path with lenses. The focal plane of the MAM was imaged by a 150× objective lens, then the image was captured with a high-resolution CMOS camera (DCC3260C, Thorlabs). The objective lens was mounted on a piezo stage (P-736.4CA, PI) with increment of 20 nm, then the distribution of focused light near the focal spot were taken by moving the piezo stage with step of 20 nm. To measure the focal efficiency, an additional aperture was placed in front of the camera with the aperture tuned to the size that we can only observe the focal spot in the camera. Afterwards, the camera was substituted by a photodiode power sensor (S142C, 350-1100 nm, Thorlabs) to measure the power at different wavelengths. The total incident light power on the MAM was measured with the same configuration, but with the aperture set to the size of the MAM image after the imaging system, and with the MAM removed.

**Imaging**

The USAF 1951 resolution target was fabricated with EBL (Raith, eLINE Plus) based patterning and Cr deposition with electron beam evaporation (Kurt J. Lesker), followed by a lift-off process. The MAM sample was placed onto the resolution target with the MAM facing down. A tape with thickness of around 100 μm was used at the edges of the glass substrate to fix the relative position between NAM and target. The image was taken by a microscope (Nikon Eclipse LV100ND) in transmission mode with a 50× NA 0.4 objective lens. By reducing the aperture size of the halogen lamp, the directivity of broadband white light source was better for imaging test. The clear imaging was obtained by tuning the focal plane of microscope away from top surface of target plane to the designed image plane. To evaluate the imaging quality at the wavelengths of 458 nm, 532 nm, and 633 nm, corresponding colour filters (FL457.9-10, FL532-10 and FL632.8-10 THORLABS) were placed on the top of the halogen lamp, respectively.

## Data Availability

The data that support the figures and other findings of this study are available from the corresponding authors upon reasonable request. Source data are provided with this paper.

## Acknowledgements

J.K.W.Y. acknowledges funding support from the National Research Foundation (NRF) Singapore, under its Competitive Research Programme award NRF-CRP20-2017-0004 and NRF Investigatorship Award NRF-NRFI06-2020-0005.

## Author Contributions

C.F.P., Hao W. and J.K.W.Y. conceived the idea of MAM. C.F. P. performed the design, numerical simulation, fabrication, and characterization with assistance from Hao W. and drafted the manuscript. Hao W. and Hongtao W. assisted in the optical characterization. P.N.S. assisted in the drawing of schematics. All the authors contributed to the data analysis and manuscript writing. J.K.W.Y. and C.-W.Q. supervised the project.

# Competing interests

The authors declare no competing interests.

# Supporting Information of Achromatic Metalenses with High Numerical Aperture Using Multilayer Structures

Cheng-Feng Pan[1,2], Hao Wang[1,3,4,*], Hongtao Wang[1,2], Parvathi Nair S[1,5], Qifeng Ruan[6], Simon Wredh[1], Yujie Ke[5], John You En Chan[1], Wang Zhang[1], Cheng-Wei Qiu[2], Joel K. W. Yang[1,*]

## Affiliations

[1] Engineering Product Development, Singapore University of Technology and Design, Singapore 487372, Singapore

[2] Department of Electrical and Computer Engineering, National University of Singapore, 4 Engineering Drive 3, Singapore, 117576, Singapore

[3] College of Mechanical and Vehicle Engineering, Hunan University, Changsha 410082, China

[4] Greater Bay Area Institute for Innovation, Hunan University, Guangzhou 511300, China

[5] Institute of Materials Research and Engineering, A*STAR (Agency for Science Technology and Research), Singapore 138634, Singapore

[6] Ministry of Industry and Information Technology Key Lab of Micro-Nano Optoelectronic Information System, Harbin Institute of Technology (Shenzhen), Shenzhen 518055, China

## Corresponding author

✼E-mail: joel_yang@sutd.edu.sg

✼E-mail: hao_wang@sutd.edu.sg

**Supplementary Note 1: Abbe number and materials dispersion**

The Abbe number is an index used to represent the dispersion ability of a transparent medium. Generally speaking, the larger the variation of refractive index of the medium, the more significant the dispersion, and the smaller the Abbe number; The general definition of the Abbe number $V_D$ of a medium is:

$$V_D = \frac{n_d - 1}{n_F - n_C}$$

where $n_C$, $n_d$ and $n_F$ are the refractive indices of the material at the wavelengths of the Fraunhofer C, d, and F spectral lines (656.3 nm, 587.56 nm, and 486.1 nm respectively). This only applies in the visible range. For more general formulations, it can be written as:

$$V = \frac{n_{center} - 1}{n_{short} - n_{long}}$$

Where $n_{center}$ is refractive index of the central wavelength of the bandwidth we are interested in, $n_{short}$ is the shortest wavelength in the bandwidth and $n_{long}$ is the longest wavelength in the bandwidth. We selected some typical materials which are used to fabricate the metalens and calculated the Abbe numbers as shown in Supplementary Figure 1.

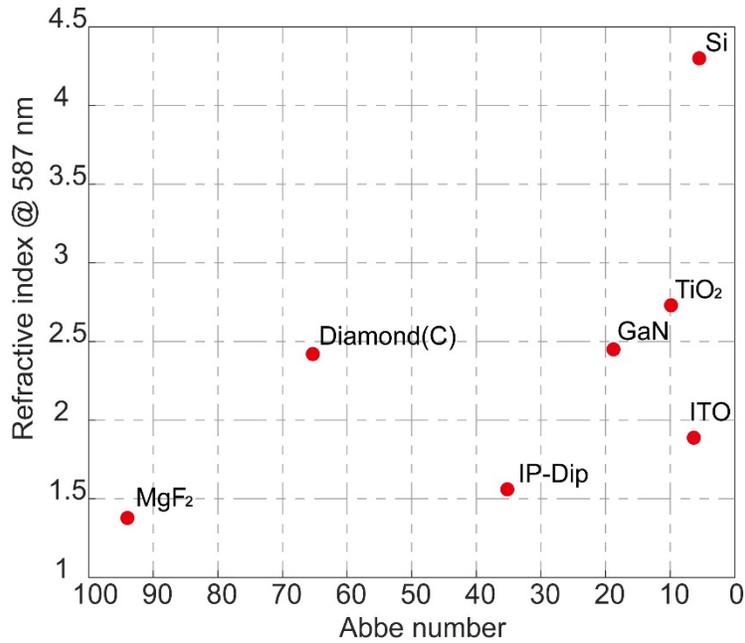

**Supplementary Figure 1.** Abbe number of some common materials used for nanofabrication from low-refractive-index (LRI) to high-refractive-index (HRI)[1, 2, 3, 4, 5, 6, 7].

We can see that high-refractive-index (HRI) materials have better abilities to manipulate light but they have more severe dispersions which will cause difficulties in designing the broadband meta-devices. Diamond has higher Abbe number but it is not suitable for common nanofabrication. In this work, IP-Dip resin is better to be used to fabricate the broadband 3D meta-device with its relative high Abbe number and ability for freefrom structure fabrication by a 3D printer with nanoscale resolution.

**Supplementary Note 2: Literature review on achromatic metalens working in the visible range**

To summarize the achromatic metalens research works in the visible range with single layer or multi-layers, we provide the following tables to show the achromatic metalens performance achieved previously. The foucsing efficiency is the average efficiency in thw band except for the ones specifying the wavelength. Most of the works utilize the HRI material to fabricate metalens, and for single layer achromatic metalenses, the NA is smaller than 0.3.

Table 1 Single layer achromatic lens

| Lens Type | Aperture (μm) | Bandwidth (nm) | NA | Material | Focusing Efficiency |
|---|---|---|---|---|---|
| Khorasaninejad et al.[8] Metalens | 200 | 490-650 | 0.2 | $TiO_2$ | 15% |
| Wang et al.[9] Metalens | 50 | 400-600 | 0.106 | GaN | 40% |
| Chen et al.[10] Metalens | 25 | 470-670 | 0.2 | $TiO_2$ | 20% at 500 nm |
| Ye et al.[11] Metalens | 7 | 435-685 | ~0.15-0.19 | Polymer | 15%-20% |
| Chen et al.[12] Metalens | 26.4 | 460-700 | 0.2 | $TiO_2$ | ~33% |
| Fan et al.[13] Metalens | 14 | 430-780 | 0.086 | SiN | 47% |
| Chung et al.[14] Metalens | 12.5 | 450-700 | 0.1 | $TiO_2$ | 65% |
| Xiong et al.[15] Micro lens | 10 | 425-700 | 0.24 | Polymer | 60% |
| Balli et al.[16] Metalens | 20 | 450-1700 | 0.27 | Polymer | 60% |
| Liu et al.[17] Metalens | 10 | 450-800 | ~0.28 | $TiO_2$ | 64% |

Table 2 Multi-layer achromatic lens

| Lens Type | Aperture (μm) | Bandwidth (nm) | NA | Material | Focusing Efficiency |
|---|---|---|---|---|---|
| Chen et al.[18] Metacorrector + Refractive lens | 1500 | 470-700 | 0.075 | $TiO_2$ + Silica | 35% |
| Yao et al.[19] Metalens | 100 | 620-675 | 0.24 | $TiO_2$+ Si | 22% |
| Mclung et al.[20] Metalens | 600 | 450-650 | 0.2 | α-Si | N.A. |
| Huang et al.[21] Metalens | 5/8.6 | 470-650 | 0.61 | TiO$_2$ | 50% |
| Li et al.[22] Metalens | 19.5 | 400-700 | 0.26 | TiO$_2$ | 50% |

**Supplementary Note 3: Upper limit of the spacing distance**

We first simulate an single layer focusing lens, the power transmitted along the $z$ axis is collected with a monitor, as shown in Supplementary Figure 2. The power attenuation after 2 μm is around 60%. In order to design a high-efficiency achromatic metalens, and to avoid the energy loss caused by the side diffraction, the spacing distance between two different layers is pre-defined to be less than 2 μm for multilayer design.

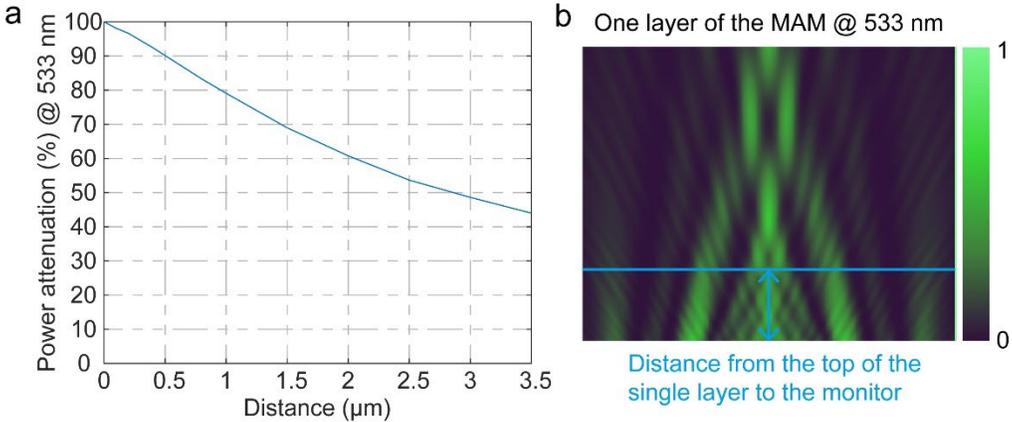

**Supplementary Figure 2**. Power attenuation along the $z$ axis. **a**. Power attenuation versus distance. **b**. The $xz$ image along the propagation plane. The distance is the length between the monitor and the top of the single layer focusing lens.

**Supplementary Note 4: Optimization process**

To make the optimization problem tractable, it is essential to make use of a rapidly converging and computationally efficient inverse design method. Here, we use the adjoint method, which enables inverse design in nano optics[23, 24], to efficiently compute gradients and accelerate convergence in topology optimization. The adjoint method involves two steps, i.e., forward and adjoint propagation. Forward propagation evaluates the intensity $|E(\xi)|^2$ at the focal point and the electric field tensor $\boldsymbol{E}_f$ in the design region with a matching tensor size as $\boldsymbol{\xi}$. Adjoint propagation obtains the same tensor size electric field $\boldsymbol{E}_{adj}$ in the design region with a dipole placed at the focal spot as the light source. To generate the adjoint gradient, the derivative of the objective over the variables can be written as, $\frac{d\Phi}{d\xi} = \frac{\partial \Phi}{\partial \xi} + \frac{\partial \Phi}{\partial E}\frac{\partial E}{\partial \xi}$. In the manuscript, we define the objective as $\Phi = |E(\xi)|^2$ and the goal is to maximize $\Phi$. For an optical system, we can use the wave equation to descripe the system with a source,

$$\mu_0^{-1} \nabla \times \nabla \times \boldsymbol{E}(\xi) - \omega_0^2 \varepsilon_0 \varepsilon_r(\xi) \boldsymbol{E}(\xi) = -i\omega_0 \boldsymbol{J}(\xi) \tag{1}$$

For the normal focusing system, it can be considered as a linear system, so Eq. (1) can be written as

$$\boldsymbol{C}\boldsymbol{E}(\xi) - b = 0 \tag{2}$$

$\boldsymbol{C}$ is a vector of constant. In order to get the gradient of the objective according to the parameters $\boldsymbol{\xi}$ we want to optimize, the chain's rule is applied,

$$\frac{d\Phi}{d\xi} = \frac{\partial \Phi}{\partial \xi} + \frac{\partial \Phi}{\partial E}\frac{\partial E}{\partial \xi} \tag{3}$$

We do a partial differentiation of Eq. (2) with respect to $\boldsymbol{\xi}$

$$\frac{\partial \Phi}{\partial \xi}\boldsymbol{E} + \boldsymbol{C}\frac{\partial \boldsymbol{E}}{\partial \xi} = 0 \tag{4.1}$$

$$\frac{\partial \boldsymbol{E}}{\partial \xi} = -(\boldsymbol{C})^{-1}\frac{\partial \Phi}{\partial \xi}\boldsymbol{E} \tag{4.2}$$

By inserting Eq. (4.2) into Eq. (3), we can get:

$$\frac{d\Phi}{d\xi} = \frac{\partial \Phi}{\partial \xi} - \left(\frac{\partial \Phi}{\partial E}(C)^{-1}\right)\frac{\partial \Phi}{\partial \xi}E \tag{5}$$

If we consider $\frac{\partial \Phi}{\partial E}(C)^{-1}$ as an adjoint field $\widehat{E}^T$, then, $\frac{\partial \Phi}{\partial E}(C)^{-1} = \widehat{E}^T$

$$\frac{d\Phi}{d\xi} = \frac{\partial \Phi}{\partial \xi} - \widehat{E}^T \frac{\partial \Phi}{\partial \xi}E \tag{6}$$

and

$$C^T\widehat{E} - \left(\frac{\partial \Phi}{\partial E}\right)^T = 0 \tag{7}$$

It fits with the wave equation with a dipole source and the source amplitude at the focal point is defined as $\frac{\partial \Phi}{\partial E} = 2 \times E$. From the chain rule and max(min) methods we applied in the main text, the amplitude of the dipole light source in $i^{th}$ round is set as $A = 2\times\min(|E(\xi_i,\omega)|)$ at the focal point. As mentioned in Eq. (2), the wavelength with the smallest focal intensity is the objective value in each iteration. Thereafter, the gradient is computed with respect to the two-step simulation results, $\Phi'(\xi) = 2 \times \text{real}(E_f \cdot E^*_{adj})$. With the assistance of the adjoint method generated gradient, the objective function can be analytically differentiated.

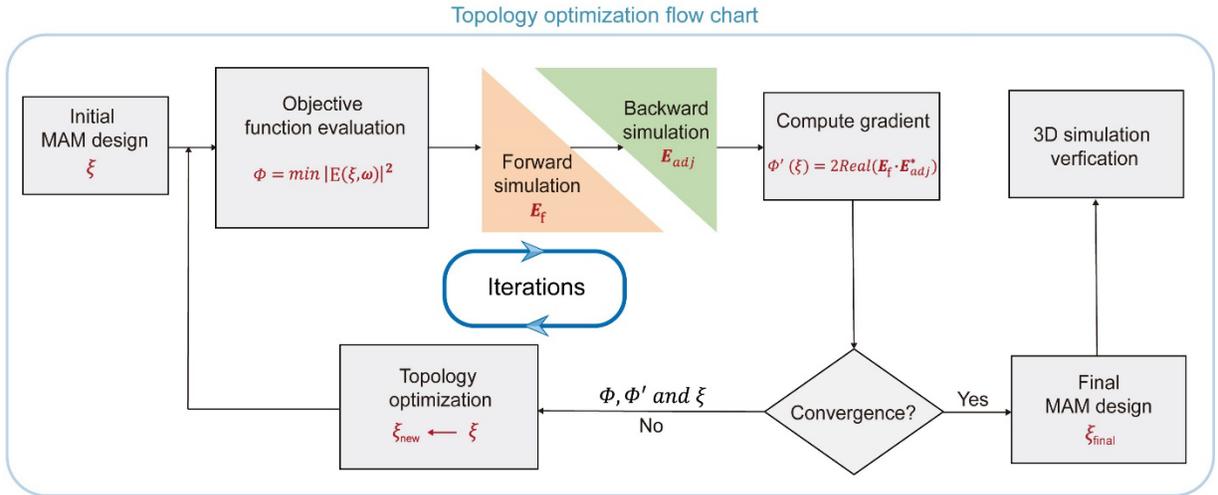

**Supplementary Figure 3**. Flow chart of the topology optimization method assisted by the adjoint method.

**Supplementary Note 5: Optimization result of single layer achromatic metalens**

With the optimization, we found that we cannot realize a 0.5 NA, 400 nm bandwidth achromatic metalens in one single layer.

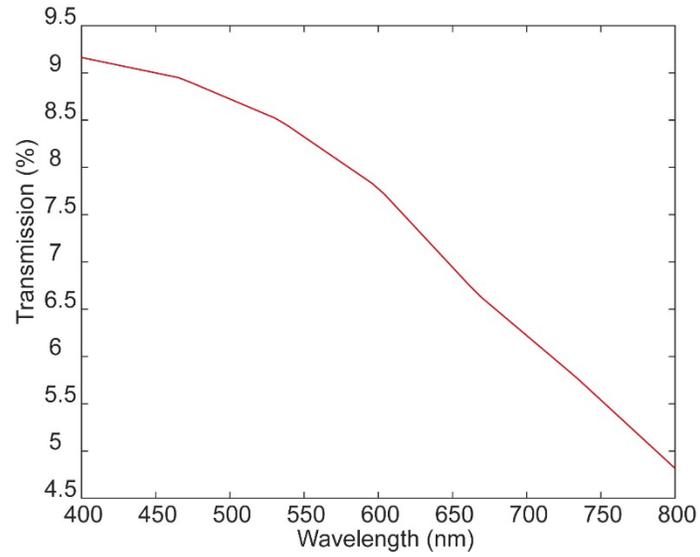

**Supplementary Figure 4**. Transmission of single layer metalens at the focal plane.

**Supplementary Note 6: 0.5 NA and 0.7 NA MAM performance in 300-1100 nm**

We simulated optical performance of 0.5 NA and 0.7 NA MAM from 300 to 1100 nm in the *xz* plane. There is only a slight displacement in *z* axis observed, which means it is possible to design MAM with wider bandwidth than the ones shown in the manuscript by the sacrifice of NA or efficiency.

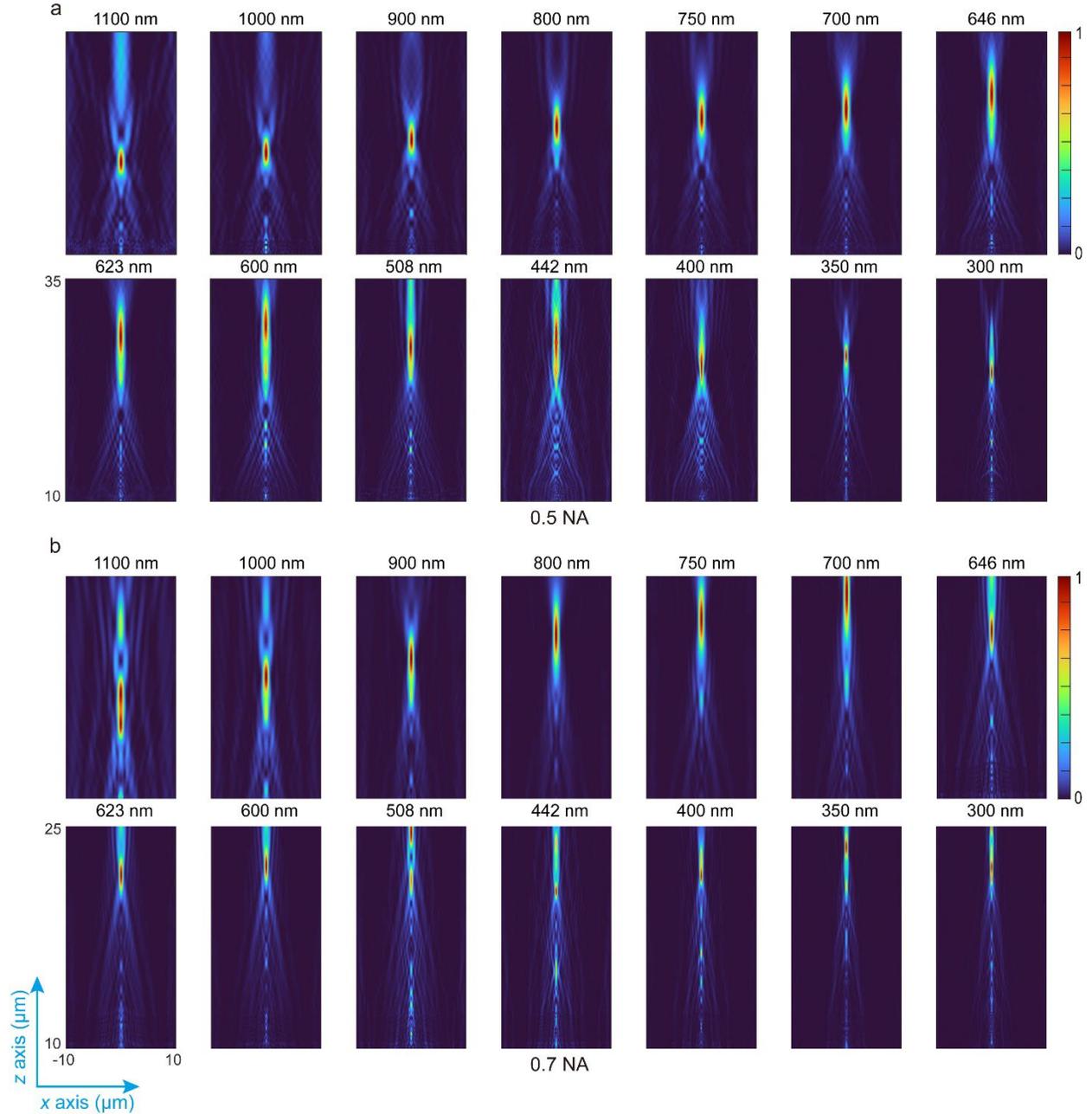

**Supplementary Figure 5**. Optical performance in the *xz* plane for **a,** 0.5 NA MAM and **b,** 0.7 NA MAM.

**Supplementary Note 7: Explanation for the outliers in Fig. 2d(iii)**

The reason for the 2 outliers is the presence of two focal points on the axis of light propagation, and the selection of the focal position based on higher intensity. Taking 509 nm as an example, we can see from the Supplementary Figure 6.

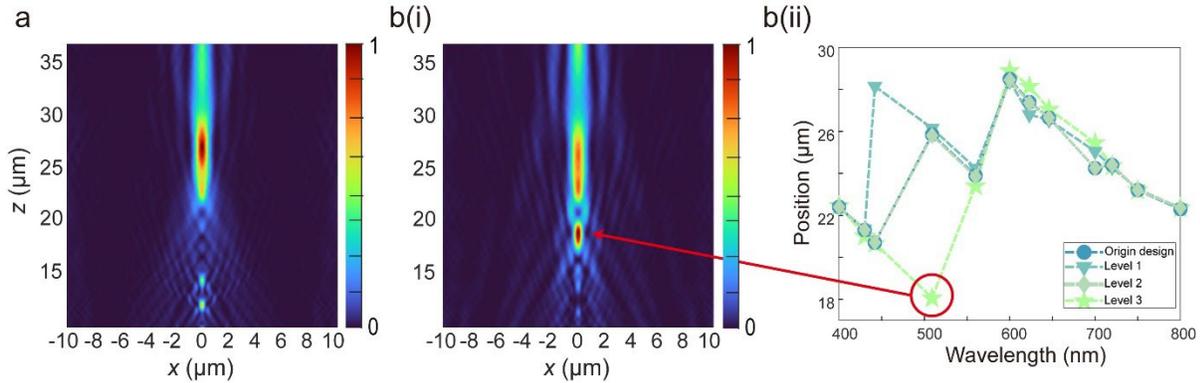

**Supplementary Figure 6**. Image of light propagation **a**. origin design and **b** (i). Level 3 design at the *xz* plane. **b** (ii). position of maximum focal intensity along the propagation axis (iii) for different levels.

**Supplementary Note 8: Point spread function and modulation transfer function of MAM**

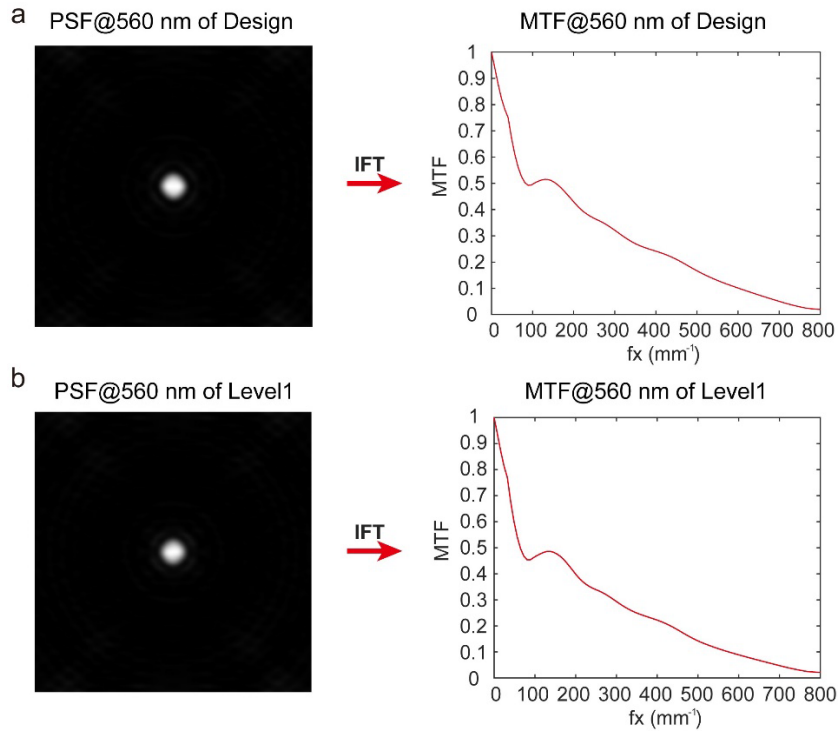

**Supplementary Figure 7.** The simulated PSF and MTF of the **a**. original design and the **b**. Level 1 design in the maintext at 560 nm.

**Supplementary Note 9: Fabrication optimization**

We carefully optimize our fabrication process with several methods. The edges of the fabricated structures are easy to delaminate from the fused silica substrate if we directly print on it. Therefore, we first spin coated the TI Prime onto the substrate to increase the adhesion between the structures and the substrate surface. Due to the limited degree of conversion of the resin during exposure process, the 3D printed structures usually suffer from shrinkage issue, which induces the deviation from design. Here, the shrinkage is solved by a precise compensation in our fabrication recipe. The detail of the compensation is shown in Supplementary Fig. 2a i-iii, and the detailed surface profile are shown in Supplementary Fig. 2b i-ii for 0.5 NA and 0.7 NA MAM.

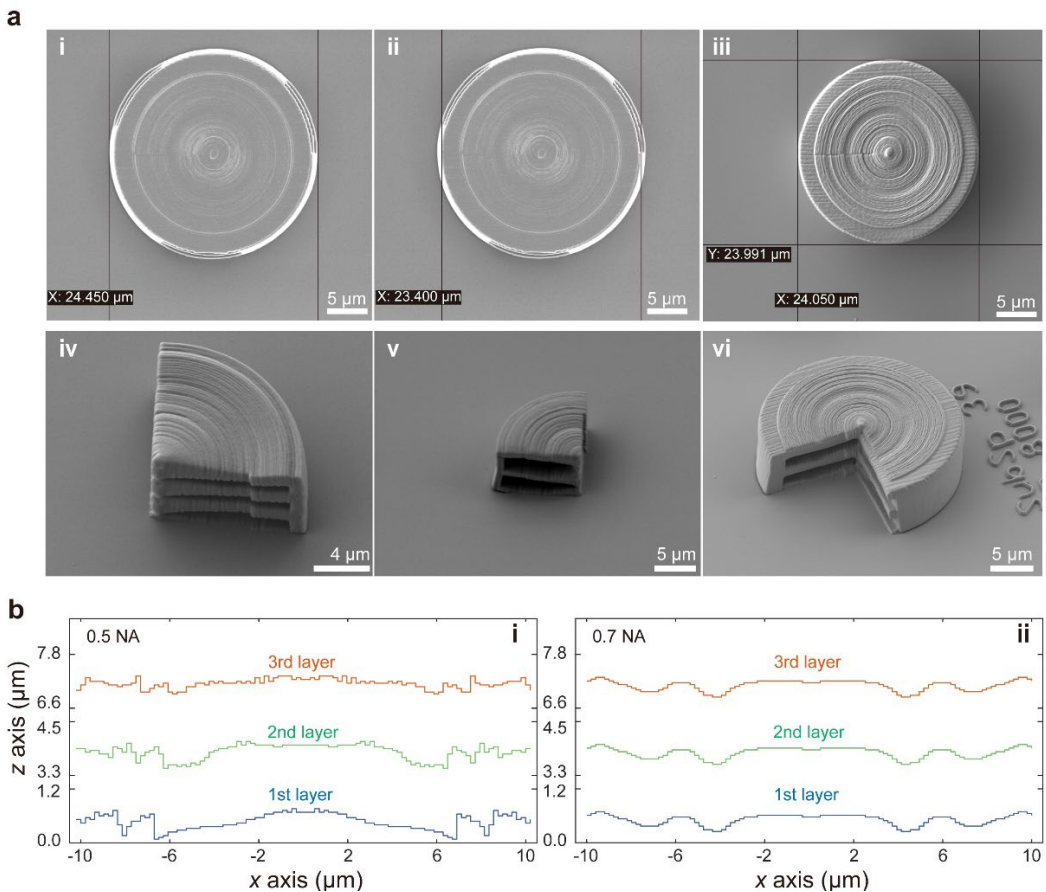

**Supplementary Figure 8.** Fabrication details of the MAM. **a,** (i-ii) SEM image of the MAM before compensation with radius along bottom $x$ axis equals to 24.45 μm, and top $x$ axis equals to 23.4 μm, and (iii) SEM image of the designed MAM structures after shrinkage compensation of radius along *x* axis and *y* axis from bottom to the top equal to 23.99 and 24.05 μm. (iv-vi) Optimization of the MAM of the adhesion and falling of layers along *z* axis. (iv) Merge effect appears between layers and between lines. (v) The falling of layers along *z* axis is solved but there is still adhesion in the center. (vi) Final optimized structrues with clearer separation between lines and layers. **b,** Detailed surface topography of 0.5 NA (i) and 0.7 NA (ii).

**Supplementary Note 10: Imaging of the resolution target**

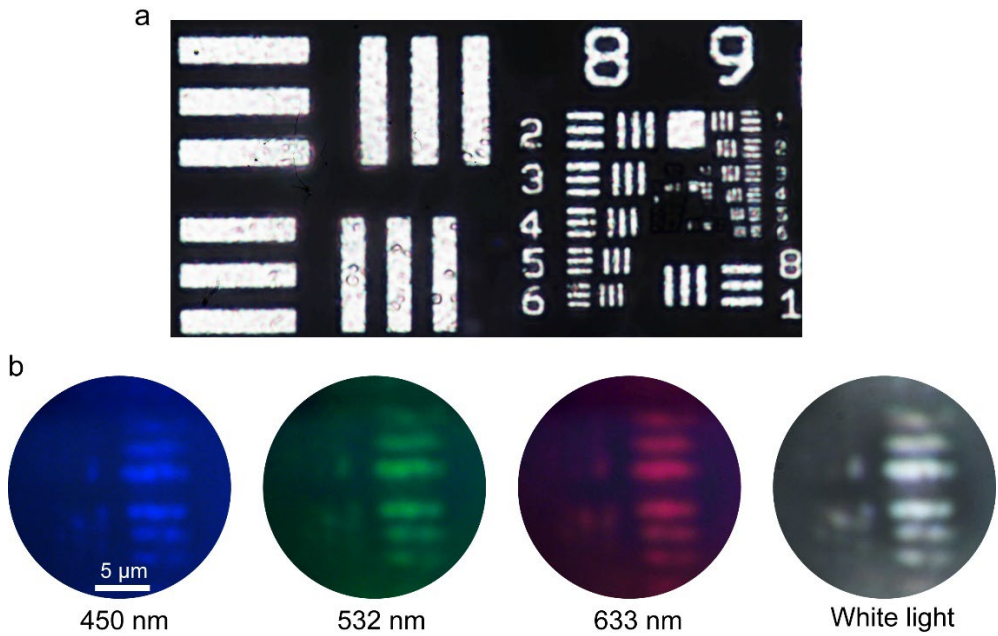

**Supplementary Figure 9. a** The microscope image of the resolution target taken by microscope in transmission mode with a 50× NA 0.4 objective lens. **b** Optical images of the number lines in group 6 element 5 and 6 in the USAF 1951 resolution target captured through the 0.5 NA MAM under white light, and applied blue (458 nm), green (532 nm), red (633 nm) illumination.

**Supplementary Note 11: 0.5 NA 3D printed spherical lens and Fresnel lens performance**

In order to compare the difference between the MAM and traditional lens in micro-scale, we 3D printed the 0.5 NA Fresnel lens and spherical lens with the same resin (IP-Dip). Obvious dispersion can be seen from these lenses in Supplementary Figure 5.

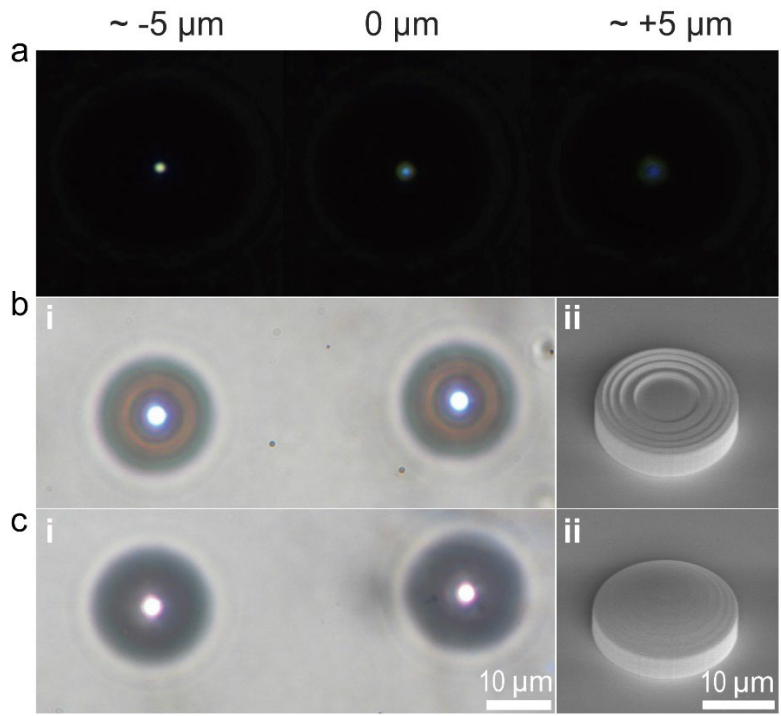

**Supplementary Figure 10**. Performance of 3D printed traditional 0.5 NA Fresnel and spherical lens. **a**, Optical microscope images of a Fresnel lens at various distances from the focal point. **b, (i-ii)** Optical microscope image of the Fresnel lens at the focal point and the SEM image of the Fresnel lens. **c, (i-ii)** Optical microscope image of the spherical lens at the focal point and the SEM image of the spherical lens.

**Supplementary Note 12: Optical testing apparatus**

The setups of the optical characterization is shown in Supplementary Fig. 3a. A supercontinuum laser light source (SuperK SELECT UV/nIR1, NKT Photonics) with wavelength range of 300–1500 nm was employed. The light path was first collimated, then the beam was illuminated on the fabricated MAM by carefully tunning the sample position. The 0.5 NA and 0.7 NA MAM were fabricated on one substrate. We then used the piezostage to control the 150× lens to focus on the focal point of the MAM and moved along the axis of light propagation to obtain the 3D power distribution of the focal spot. For the detection of focusing efficiency, the CMOS (DCC3260C, Thorlabs) was susbititued by a power detector (S142C, 350-1100 nm, Thorlabs). The schematic of the imaging setup is show in Supplementary Fig. 6b.

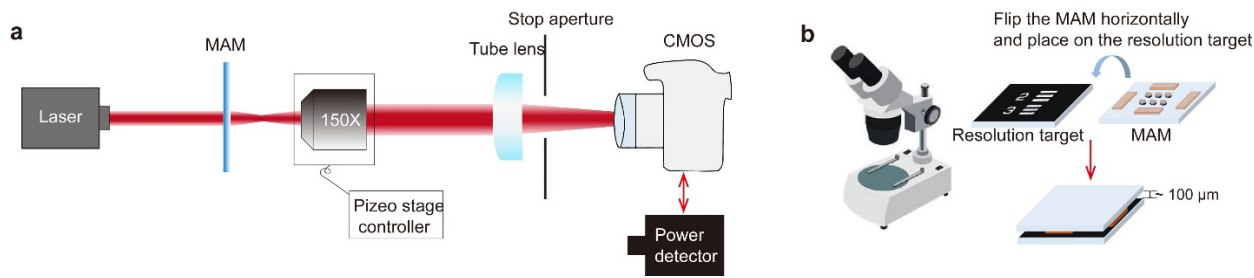

**Supplementary Figure 11.** Optical characterization and imaging testing setup. **a,** Optical testing setup. After the light propagating through the MAM, the focal spot is imaged by a 4f system with a 150× lens and a tubed lens and the image is finally collected by a CMOS camera. The stop aperture is used to restrict the area of light collected by the power detector. **b,** The MAM imaging setup.